\documentclass[prd, aps, superscriptaddress, preprintnumbers, twocolumn, floatfix, nofootinbib]{revtex4}

\usepackage{amsfonts}
\usepackage{amsmath}
\usepackage{amssymb}
\usepackage{bm}
\usepackage{dcolumn}
\usepackage{graphicx}   
\usepackage[latin1]{inputenc}
\usepackage{latexsym}
\usepackage{rotating}
\usepackage{hyperref}
\usepackage{graphicx}
\usepackage{color}

\newcommand\be{\begin{equation}}
\newcommand\ba{\begin{eqnarray}}
\newcommand\ee{\end{equation}}
\newcommand\ea{\end{eqnarray}}

\begin{document}

\title {Unified Dark Matter, Dark Energy and Baryogenesis via a ``Cosmological Wetting Transition''}

\author{Robert Brandenberger}
\email{rhb@physics.mcgill.ca}
\affiliation{Department of Physics, McGill University, Montr\'{e}al, QC, H3A 2T8, Canada and
Institute of Theoretical Physics, ETH Z\"urich, CH-8093 Z\"urich, Switzerland}

\author{J\"urg Fr\"ohlich}
\email{juerg@phys.ethz.ch}
\affiliation{Institute of Theoretical Physics, ETH Z\"urich, CH-8093 Z\"urich, Switzerland}

\author{Ryo Namba}
\email{namba@physics.mcgill.ca}
\affiliation{Department of Physics, McGill University, Montr\'{e}al, QC, H3A 2T8, Canada}

\date{\today}

\begin{abstract}

In a recent publication \cite{us}, a cosmological scenario featuring a scalar field, $\varphi$, that is a source for Dark Matter and Dark Energy has been proposed. In this paper, a concrete realization of that scenario is presented. As in many models of scalar-field driven Dark Energy, the effective Lagrangian of the field $\varphi$ contains a potential proportional to $e^{-\varphi/f}$. This potential is modulated in such a way that, in the absence of other matter fields, it has a local minimum at a small value of $\varphi$. Fluctuations of $\varphi$ around this minimum give rise to a gas of dark-matter particles. The field $\varphi$ is coupled to another scalar field $\chi$ in such a way that the minimum in the effective potential of $\varphi$ disappears when, after a continuous phase transition accompanied by spontaneous symmetry breaking, $\chi$ develops a non-vanishing expectation value. This triggers slow growth of a homogeneous component of $\varphi$ accompanied by the emergence of Dark Energy, a phenomenon analogous to the ``wetting transition'' in statistical mechanics. Inside regions of the Universe where the pressure is small and the energy density is large enough to stall expansion, in particular around galaxies and galaxy clusters, the phase transition in the state of $\chi$ does not take place, and a gas of cold dark-matter particles persists. The evolution of $\varphi$ at very early times may tune the rate of baryogenesis.

\end{abstract}

\maketitle

\section{Introduction} 
\label{sec:intro}

In a recent paper \cite{us} we have proposed a scenario incorporating a scalar field,
$\varphi$, that yields a unified description of Dark Matter and Dark Energy. The idea underlying our proposal
is that, at early times in the evolution of the Universe, $\varphi$ is trapped in the vicinity
of a homogenoeus configuration $\varphi_0$, and oscillations of $\varphi$ around $\varphi_0$ 
form a gas of dark-matter particles.
It is envisaged in \cite{us} that, at a late time $t_c$, a phase transition takes place allowing a
homogeneous component of $\varphi$ to emerge that gives rise to dynamical Dark Energy. The 
transition time $t_c$ has to be tuned to be later than the time of onset of cosmological structure 
formation, but earlier than the present time.
In regions of the Universe where the energy density is high but the pressure is tiny the 
expansion of space is stalled, the phase transition does not take place, 
and Dark Matter described by oscillations of $\varphi$ continues to prevail. 
Thus, at late times close to the present time, a single scalar field, namely $\varphi$, 
is a source for Dark Matter inside overdense regions of the Universe, but a source 
for Dark Energy on cosmological scales. 

The purpose of this paper is to describe a concrete realization of the scenario sketched in \cite{us}, drawing some
inspiration from the phenomenon of the ``wetting transition'' in statistical mechanics \cite{Wetting}. 
The field $\varphi$, which has a modulated effective
potential decaying exponentially at large field values, as originally proposed in \cite{quint},
is coupled to a second scalar field, $\chi$, with a 
quartic potential that can lead to a low-temperature phase transition accompanied by the 
spontaneous breaking of a continuous symmetry. We denote the transition temperature by $T_c$.
At temperatures above $T_c$ and/or small values of $\varphi$, the field $\chi$ is trapped near its symmetric
configuration, $\chi = 0$. At very early times, when the temperature, $T$, of the Universe is very high, 
the effective potential of $\varphi$ does not have a local minimum, and the expectation 
value of $\varphi$ increases. But as $T$  decreases, yet $T>T_c$, 
the effective potential of $\varphi$ develops a local minimum at a small field value, $\varphi_0$. 
Oscillations of $\varphi$ around the configuration $\varphi=\varphi_0$ give rise to Dark
Matter. When the temperature of the background cosmology drops below $T_c$ (at a time $t_c$)
the quantum state of the Universe has the property that $\chi$ has a non-zero expectation value.\footnote{This 
signals the breaking of a continuous symmetry. Whether this process is accompanied by the emergence of
a massless Goldstone mode in the particle spectrum, or not, depends on whether the broken symmetry is global or local, 
(in which case $\chi$ is coupled to a gauge field).} After this transition, the local minimum in the effective potential of $\varphi$ disappears again, and a spatially homogeneous, slowly growing configuration of the field $\varphi$ develops  
that gives rise to Dark Energy. The phase transition does, however, not take place inside
``overdense'' regions, i.e., in the vicinity of galaxies and galaxy clusters, and, in such regions, $\varphi$
continues to oscillate around a small constant value, hence giving rise to Dark Matter. In conclusion, our scenario may
provide a unified description of Dark Matter and Dark Energy. 

The field $\varphi$ might actually be of importance in yet another context: If one couples 
the gradient $\partial_{\mu} \varphi$ linearly
to the baryon current, $j^{\mu}_{B}$, (thus violating CP) then, at early times 
(before $\varphi$ starts to oscillate around $\varphi_0$), $\dot{\varphi}$ has the function
 of a ``chemical potential'' tuning the matter-antimatter asymmetry during baryogenesis 
\cite{Cohen}; see also \cite{other}.\footnote{The idea that $\dot{\varphi}$ may represent a kind of
``chemical potential'' conjugate to the baryon number is similar to the idea that the time 
derivative of a pseudo-scalar axion may act as a chemical potential 
tuning the difference of left-handed and right-handed light leptons, hence giving rise to the ``chiral magnetic effect''
\cite{Vilenkin}, and, thus, it may play a significant role in understanding the origin of primordial 
magnetic fields in the Universe \cite{magn-fields}. Helical (hyper) magnetic fields may play and important role in baryogenesis;
see \cite{Kamada}.}
 
The model discussed in this paper has various advantages over other models of
unified Dark Matter and Dark Energy; (see \cite{unified} for a review of early work
on unified models): Since the square of the speed of sound, $c_s^2$,
is negligibly small at early times ($t < t_c$), structure formation proceeds 
as in the standard $\Lambda$CDM model up to the time $t_c$ when the phase transition takes place,
and since, after the transition, $c_s^2 = 1$ in the bulk, no dangerous instabilities for fluctuations at late
times arise, as they do in certain quartessence models \cite{quart}; (see the
discussion in \cite{Rodrigo}).

Our model ought to be viewed as a toy model capturing some of the features one would have to require of a unified theory of Dark Matter and Dark Energy. In a future paper, we will present a more systematic survey of different such models.

Note that there have been earlier works proposing that Dark Matter and Dark Energy emerge from the same scalar field. In particular, Wetterich proposed a {\it cosmon dark matter} model \cite{Wett2} in which dark matter is postulated to be fluctuations of the same setup \cite{Wett1}l in which the background scalar field represents Dark Energy.  Our realization of the unified Dark Matter and Dark Energy Scenario is, however, very different.
 
We will use natural units in which the speed of light, Planck's constant and Boltzmann's
constant are set to 1. The cosmological scale factor is denoted by $a(t)$, the
Hubble expansion rate by $H(t)$, and the Planck mass by $m_{pl}$. Temperature
is denoted by $T$, and $p$ and $\rho$ stand for pressure and energy density,
respectively.

\section{The Model} \label{model}

We consider an effective field theory featuring two (for concreteness canonically normalized) scalar fields,
 $\varphi$ and $\chi$, with a potential
\be \label{bare}
V(\varphi, \chi) \, = \, \bigl( M^2\, \varphi^2 + \mu^2 \vert \chi \vert^{2} + \epsilon_0 \bigr) e^{- \varphi/f} 
 + \frac{\lambda}{4} \bigl( \vert \chi \vert^{2} - \eta^2 \bigr)^2 \, ,
\ee
where $\lambda > 0$ is a dimensionless coupling constant,
$\eta$ is related to the vacuum expectation value of $\chi$, $M$ is a positive constant with
(mass) dimension $1$, with $M^2\, \varphi^{2}$ modulating the exponential potential of $\varphi$,
$\mu$ is a constant of dimension $1$ describing the coupling
of $\chi$ to $\varphi$, and $\epsilon_0$ is a (not necessarily positive) constant
of dimension $4$. (The term $M^2\,\varphi^{2}$ might be generated by couplings of $\varphi$ to other, very massive degrees of freedom not explicitly considered here.) The terms in the effective potential that depend explicitly on $\varphi$ are 
proportional to $\text{exp}[-\varphi/f]$, a behavior used in many quintessence 
models of Dark Energy \cite{quint}. The constant $f$ defines the field range above which the potential
decays.  As we will see later, values of the paramters which satisfy the constraints on our model are $f \sim m_{pl}$, $\mu \sim \eta \sim 10^{-2} {\rm eV}$ (up to a coupling constant $\lambda$) and $M \sim 10^{-28} {\rm eV}$. We will in fact be setting $\epsilon_0 = 0$.

There are tight constraints \cite{Carroll} on the couplings of a dark-energy field to Standard-Model degrees of freedom. In the following, we usually assume that $\varphi$ is not coupled directly to any fields appearing in the Standard Model (see, however, Sect. V). It is natural to assume that $\varphi$ has a geometrical origin. It may be related to the radius of an extra (spatial) dimension, as in {\it Horava-Witten} theory \cite{HW}, see also
\cite{CFG}, or in the {\it Brane World} scenario \cite{brane}:
\be
\varphi \, = \varphi_0  {\rm{ln}} \left(\frac{r}{\ell_s} \right) \, ,
\ee
where $r$ is the (space-time dependent) radius of an extra dimension, and $\ell_s$ is a fundamental length -- 
for example the string length if we have in mind an effective field theory coming from string theory -- and $\varphi_0$ is a normalization constant. In this context, it is natural to choose the scale $f$ to be proportional to the four-dimensional Planck mass.

We will argue below that if our model yields a realistic cosmological scenario then, at early times in the evolution of the Universe, the expectation value of $\chi$ would have to have vanished. The constants appearing in the effective potential $V$ of Eq. \eqref{bare} are chosen such that, as a function of $\varphi$, $V$ has a local minimum as long as $\vert \chi \vert^{2}$ is small enough. However, for large values of $\vert \chi \vert^{2}$, $V$ does not have any local minimum in $\varphi$. Specifically, setting $\chi = 0$ in \eqref{bare}, the potential has 
extrema at values of $\varphi$ given by
\be\label{crit-pts}
\varphi_{\pm} \, = \, f \pm \sqrt{f^2 - \frac{\epsilon_0}{M^2}} \, .
\ee
The right side of \eqref{crit-pts} is real-valued, provided
\be\label{lowerbound}
f^2 \, > \, \frac{\epsilon_0}{M^2} \, .
\ee
We will choose $\epsilon_0$ to cancel the potential energy coming from the $\varphi$-field at the local minimum of the potential, i.e.,
\be \label{cond1}
M^2 \varphi_{-}^2 \, + \, \epsilon_0  \, = \, 0 \, .
\ee
Now (\ref{crit-pts}) becomes an equation for $\varphi_{-}$, with solution $\varphi_{-} = 0$, and the local maximum is \mbox{$\varphi_{+} = 2 f$}, which in turn sets $\epsilon_0 = 0$.

However, after the phase transition, i.e., for 
\be
\vert \chi \vert^{2} \underset{\sim}{>} \eta^{2} \, ,
\ee
we want the local minimum in the $\varphi$-direction of the potential $V$ to disappear. Given the
constraint (\ref{cond1}), the condition on the parameters for this to hold true is given by
\be \label{cond2}
f^2 \, < \, \frac{\mu^2 \eta^2 + \epsilon_0}{M^2} \, = \, \frac{\mu^2 \eta^2}{M^2} \, ,
\ee
where in the last step we have inserted the special value of $\epsilon_0$ used.

Figure 1 shows a plot of the potential $V$ as a function of $\varphi$, for $\chi = 0$ and for
$\vert \chi \vert^{2 }= \eta^{2}$. The values of the parameters $f, M, \mu, \lambda$ and $\varepsilon_{0}$ used in Figure 1 are indicated there. (They are not the values used to construct a realistic cosmology, but they yield graphs that display some key features in a clear fashion.)

\begin{figure}[h!]
  \includegraphics[width=\hsize]{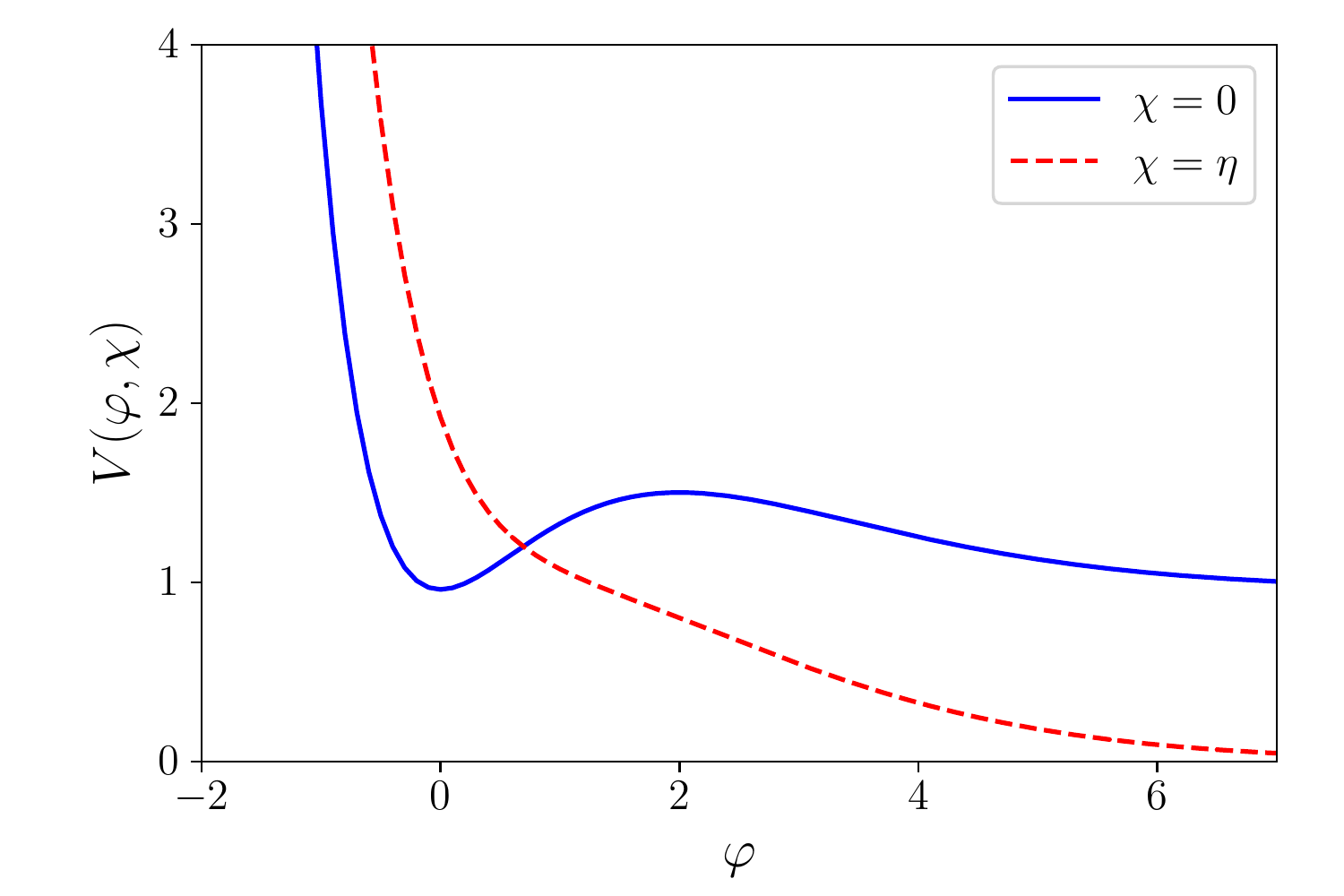}
\caption{The potential $V(\varphi, \chi)$ of our model for $\chi = 0$ and
$\chi = \eta$. The horizontal axis is $\varphi$ in units of $f$, the vertical axis is the
value of the potential in units of $f^2 M^2$. The values of the constants
were taken to be $\eta = 1.4$ in units of $\sqrt{f M}$, $\lambda = 1$ and $\epsilon_0 = 0$. The value of
$\mu$ is fixed by (\ref{cond5}). }
\label{Fig1}
\end{figure}

When the field $\chi$ is in thermal equilibrium at a temperatures $T \gg T_c$, the $\chi$-dependent terms in the potential $V(\varphi, \chi)$ given in Eq. \eqref{bare} yield temperature-dependent corrections to the effective potential, the leading ones being given by
\be \label{corr}
\Delta V(\varphi, \chi) \, = \, \langle \vert \chi \vert^{2} \rangle_{T} \left(  \lambda \vert \chi \vert^{2} -\frac{\lambda}{2}\eta^{2}+ \mu^{2} e^{-\varphi/f}\right)\, ,
\ee
where the term in angular parentheses is the expectation value of  $\vert \chi \vert^{2}$ in an equilibrium state at temperature $T$. Dimensional analysis suggests that
\be
\langle \vert \chi \vert^{2} \rangle_{T} \, \sim \, O(T^2) \, ,
\ee
and, as a consequence of red-shifting, a formula like this continues to hold after $\chi$ has decoupled from radiation. Let $T_i$ be the temperature of radiation at a time when $\chi$ was in a state of thermal equilibrium with radiation and equipartition of energy held. If thermal equilibrium is established by a quartic $\chi$ self-interaction, and if the $\chi$ distribution is close to Gaussian, then we have that
\be \label{InCond}
\lambda \langle \vert \chi \vert^{2} \rangle_{T_i}^{2} \, = \,C^{2} \, T_i^4\,,
\ee
for some dimensionless constant $C = O(1)$ which is in principle determined by
details of thermal processes but is left as a free parameter for our study. Hence, using that
\begin{equation}\label{redshift}
\langle \vert \chi \vert^{2} \rangle_{T}\, \simeq \,\frac{T^{2}}{T_{i}^{2}} \langle \vert \chi \vert^{2} \rangle_{T_i}\,,
\end{equation}
we find that
\ba\label{corr2}
\Delta V \, &=& \lambda^{-1/2} C T^{2}\left(\lambda \vert \chi \vert^{2} -\frac{\lambda}{2}\eta^{2} + \mu^{2} e^{-\varphi/f}\right)\, \nonumber\\
&=& \frac{C\sqrt \lambda}{2} T^{2}\big(2\vert \chi \vert^{2} - \eta^{2}\big)
+ \frac{C\mu^{2} T^{2}}{\sqrt \lambda} e^{-\varphi/f}\,.
\ea
The key point is that, at early times when the temperature is high ($T>T_c$), the correction term (\ref{corr}) to the effective potential keeps the field $\chi$ trapped near $\chi = 0$; but at some critical temperature $T_c$, determined by
\be \label{crit}
\lambda \langle \vert \chi \vert^{2} \rangle_{T_c} \, \simeq \, \frac{\lambda}{2} \eta^2 - \mu^2 e^{- \varphi_i  / f} \, ,
\ee
where $\varphi_i$ is the value of $\varphi$ at the time $t_c$ corresponding to $T_c$ (and, as we
have argued, $\varphi_i \sim \varphi_{-} = 0$), a symmetry-breaking phase transition occurs, and the expectation value of $\chi$ rolls towards $\eta$.\footnote{In order to avoid
the production of domain walls, $\chi$ must be a multi-component scalar field with a continuous symmetry.}

In voids, the expectation value of $\chi^2$ redshifts according to the law in Eq. \eqref{redshift}, 
and condition \eqref{crit} becomes
\be \label{T_c}
 C\sqrt{\lambda} T_{c}^{2} \, = \, \frac{\lambda}{2}\eta^2 - \mu^{2} e^{- \varphi_i  / f} \, .
\ee
However, in regions of the Universe where the energy density is comparatively large, 
which have decoupled from the Hubble flow, i.e., in the vicinity of galaxies and galaxy clusters, 
the expectation value of $\vert \chi \vert^{2}$ ceases to decrease and does not redshift, any longer.
We must demand that the phase transition happen after density perturbations
on galaxy- and galaxy-cluster scales have gone non-linear. If this is the case then $\chi$ does not undergo a symmetry-breaking phase transition in these regions, and the oscillations of $\varphi$ continue to describe Dark Matter. Note that, since $\lambda \eta^2 \gg H^2$, the phase transition takes place rapidly.

Eq. (\ref{crit}) shows that a necessary condition for a phase transition to
occur is
\be
2 \mu^2 e^{- \varphi_i / f} \, < \, \lambda \eta^2 \, .
\ee
Since before the phase transition $|\varphi| <  f$, this condition becomes
\be \label{cond3}
 \mu^2 \, < \, c \lambda \eta^2 \, ,
\ee
for some constant $c$ of order $1$.

The temperature-dependent correction terms (\ref{corr}) in the effective potential of the fields $\varphi$ and $\chi$ affect the conditions to be imposed for our scenario to work. The condition that, for $\chi = 0$, a local minimum of the effective potential in the $\varphi$-direction exist at a temperature $T$ becomes
\be \label{cond1A}
f^2 M^2 \, > \,  \epsilon_0 + \mu^2 \langle \vert \chi \vert^{2} \rangle_T  \, .
\ee
We need this condition to be satisfied at temperatures substantially higher than $T_c$, otherwise the local minimum in the potential for $\varphi$ does not exist long enough, and oscillations of $\varphi$ about this minimum (oscillations that form the Dark Matter) are not established. We will denote the temperature where the minimum of the potential appears by $T_m$ with $T_m \gg T_c$. Since the second term on the right hand side of (\ref{cond1A}) is larger than the absolute value of the first term, (\ref{cond1A}) becomes
\be \label{cond1B}
f^2 M^2 \, > \,  \mu^2 \langle \vert \chi \vert^{2} \rangle_{T_m} \, .
\ee

As discussed before, the condition that the local minimum of the effective potential in the $\varphi$-direction disappear after the phase transition, i.e., for $\vert \chi \vert \approx \eta$, implies that the extra term appearing in the expression for the prefactor of the exponential, $\exp(- \varphi/f)$, in the effective potential must be larger than $f^2$. This yields the constraint that
\be \label{cond2}
\mu^2 \eta^2 \, > \, f^2 M^2 \, .
\ee

The lower bound (\ref{cond1B}) and the upper bound (\ref{cond2}) are consistent provided that
\be \label{range}
\eta^2 \,\,\,  > \,\,\, \langle \vert \chi \vert^{2} \rangle_{T_m} \, \sim \, \lambda^{-1/2} C T_m^2 \, 
\ee
and $f^2 M^2 / \mu^2$ lies between the two bounds.

In order to satisfy the condition that the phase transition in the state of the field $\chi$ take place at very low temperature $T_c$ we will assume that the two terms in (\ref{T_c}) are both large in magnitude but cancel up to the small contribution $C\sqrt{\lambda} T_{c}^{2}$ . This means that we enforce the condition
\be \label{cond5}
\mu^2 \, = \, \frac{1}{2} \lambda \eta^2 \, .
\ee

\section{The Cosmological Scenario}

In this section we sketch a cosmological scenario of which we expect that it can be derived from our model. 
We assume that the state of the very early Universe has the properties that the symmetry 
of the $\chi$-sub-theory is unbroken, and the field $\varphi$ is homogeneous with a large negative value. Then
$\chi$ oscillates around $\chi = 0$, while $\varphi$ slides down the potential hill towards
the minimum of its effective potential at $\varphi \sim \varphi_{-} = 0$. We suppose that, after inflation,
the field $\varphi$ does not dominate the energy density of the Universe, rather that radiation
dominates and determines the time evolution of the Hubble expansion rate $H$. (We will check
the conditions on the parameters in our Lagrangian required to justify this assumption.)

Next, we argue that the model considered in this paper may yield a unified description of Dark Matter and Dark Energy related to the evolution of the field $\varphi$. Our choice of an effective potential for the fields 
$\varphi$ and $\chi$ displayed in Eq. \eqref{bare} shows that, as long as the field $\chi$ is trapped near $\chi=0$, its contribution to the effective potential of $\varphi$ is small, and, as long as condition (\ref{cond1A}) is satisfied, the potential has a minimum at
$\varphi = \varphi_{-} \approx 0$; see \eqref{crit-pts}.
Once the field $\varphi$ approaches this minimum, it starts to oscillate around it. 
These oscillations act as Dark Matter.\footnote{If, at early times, the contribution of
$\varphi$ to the total energy density is small enough, then the kinetic energy of this field, once it approaches the 
minimum of its effective potential, is too small for it to ``overshoot'' the potential barrier and continue to 
slowly grow towards $\infty$. Thus, we do not have to worry about an
``overshooting problem'' \cite{overshoot}.} The square of the mass of the dark-matter particles is given by twice the 
coefficient of the quadratic term in a Taylor expansion of the effective potential of $\varphi$ 
at its minimum, $\varphi_{-}$. Since, before its symmetry-breaking phase transition and at moderate temperatures of the Universe, the field $\chi$ can be set to zero in the calculation of the effective potential of $\varphi$, 
the square of the mass of dark-matter particles is thus given by
\ba
\frac{\partial^2 V(\varphi, 0)}{\partial \varphi^2} \bigg\vert_{\varphi=\varphi_{-}} \!\!\! &=&
\left( 2M^2 - 4M^2\frac{\varphi_{-}}{f} + \frac{M^2}{f^2} \varphi_{-}^2  + \frac{\epsilon_0}{f^2}
\right) e^{-\frac{\varphi_{-}}{f}}\nonumber \\
&\simeq& \,  2 M^2\, ,
\ea
In the second line we have used Eqs. \eqref{crit-pts} - \eqref{cond1} and the fact that 
$\varphi_{-} = \epsilon_{0} = 0$. Thus, the effective mass of dark-matter particles is given by
\be\label{DM-mass}
m_{DM} \, \simeq \, \sqrt{2} M \, .
\ee
When $\varphi$ slides down its potential hill, approaching the minimum of its effective potential, 
the Universe continues to expand and cool, eventually approaching the transition between its 
radiation-dominated early phase and a phase dominated by Dark Matter (at a time denoted by $t_{eq}$). 

The field $\chi$ keeps oscillating around $\chi = 0$ until, at a later time $t_c$ when the background temperature $T$ equals the temperature $T_c$ of the $\chi$-phase transition, it starts to develop a non-vanishing expectation value.
This transition is accompanied by the disappearance of the minimum in the effective potential of the field $\varphi$,
a phenomenon analogous to the ``wetting transition'' in statistical mechanics \cite{Wetting}.
A homogeneous component of $\varphi$ then develops that grows slowly (logarithmically) towards $\varphi=\infty$. In this late phase, the field $\varphi$ describes Dark Energy.\footnote{The field $\chi$ is not expected to contribute much to Dark Energy, because, after the phase transition, its vacuum energy vanishes.} 
A necessary condition for the phase transition in the state of the field $\chi$ to occur is (\ref{cond1B}), and demanding that this phase transition take place when $\vert \varphi \vert = 0$ then yields the condition (\ref{cond3}) for the value of $\eta$.

The phase transition in the state of the field $\chi$ does, however, \textit{not} take place in regions of the Universe that have decoupled from the Hubble flow, where the effective temperature remains higher than $T_{c}$. (In a  self-consistent analysis, this effect is seen to be due to the higher temperature of the degrees of freedom described by $\chi$ 
in ``overdense'' regions, on one hand, and to the term $\mu^2 \vert \chi \vert^{2} e^{- \varphi/f}$
in the effective potential $V(\varphi, \chi)$, see \eqref{bare}, on the other hand, which helps confining
the field $\chi$ to a vicinity of $\chi=0$, as long as $\varphi$ remains small enough, which is expected to be the case in ``overdense'' regions.) Hence, if we choose parameters in our model such that the time $t_c$ of the phase transition in the state of $\chi$ is later than the time when proto-galaxies and proto-clusters decouple from the Hubble flow then the effective potential of $\varphi$ in such ``overdense'' regions still has a minimum, and oscillations of $\varphi$ around this minimum continue to act as Dark Matter. 

\section{Parameter Values}

In this section we discuss constraints on the values of the parameters appearing in
the Lagrangian of our model derived from the requirement that the model yield a realistic cosmology.
The Lagrangian contains the following parameters: the field range $f$, the energy
scale $\eta$, the self-coupling constant $\lambda$, the parameter $\epsilon_0$,
and the mass scales $\mu$ and $M$. We attempt to identify a region of values of these 
parameters that make our model compatible with observational data.

In order for $\varphi$ to give rise to Dark Energy today, $\varphi$ must grow slowly at the present time. This implies that %
\be
f \, \geq \, m_{pl} \,,
\ee
see \cite{exppot}.
To avoid introducing a new mass hierarchy we assume that $f = m_{pl}$, in
the following.

In Sect. II, we have found several additional conditions. The first one is that the effective potential of $\varphi$ have a minimum at times corresponding to a range of temperatures above the transition temperature $T_c$. This leads to (\ref{cond1B}). The second condition is that the minimum disappear after the phase transition, when $|\chi| \propto \eta$. This leads to the constraint in (\ref{cond2}). These two conditions are self-consistent provided that (\ref{range}) is obeyed. To obtain the phase transition at a sufficiently low temperature $T_c$, the condition (\ref{cond5}) needs to be imposed.

Another important requirement on the parameters of the model is that, 
at the end of the dark-matter phase, the energy density corresponding to the minimum of the
effective potential does not dominate the energy density. Focusing on the energy density in the $\chi$-field yields the constraint
\be \label{cond6}
\lambda \eta^4 \, \leq \, g^* T_c^4 \frac{T_{eq}}{T_c} \, ,
\ee
where $g^*$ is the number of radiative helicity degrees of freedom,
and the last factor on the right side comes from the
fact that, for $t< t_{eq}$ (with $t_{eq}$ the time of equal matter
and radiation), the total energy density is larger than the energy density of radiation
by that factor. 

The $\chi$-phase transition can only occur if the resulting increase
in the potential energy of $\varphi$ does not exceed the
energy released by $\chi$. An equivalent way of phrasing
this condition is that, at the beginning of the dark-energy phase, the potential energy
of $\varphi$ cannot exceed the energy of matter. Using the
value $\varphi \approx f$ for the field $\varphi$ right after the transition the
condition becomes
\be \label{cond7}
M^2 f^2 + \epsilon_0 \, < \, g^{*} T_c^4 \frac{T_{eq}}{T_c} \, ,
\ee
which is satisfied if $\epsilon_0 = 0$ is chosen as in Section II and (\ref{cond2}), (\ref{cond5}) and (\ref{cond6}) are used.

From now on, we choose $T_c = 10^{-1} {\rm{eV}}$. Then, (\ref{cond6}) will fix the value of $\eta$, once the coupling constant $\lambda$ is chosen. We have that
\be \label{etavalue}
\eta \, \approx \lambda^{-1/4} (g^*)^{1/4} T_{c} \left( \frac{T_{eq}}{T_c} \right)^{1/4} \, .
\ee
The value of $\mu$ is then determined by (\ref{cond5}), which yields
\be
\mu \, \approx \, \frac{1}{\sqrt{2}} \lambda^{1/4} (g^*)^{1/4} T_{c} \left( \frac{T_{eq}}{T_c} \right)^{1/4} \,.
\ee

The conditions for the existence and disappearance of the local minimum of the effective potential (discussed in Section II - see (\ref{range})) yield an upper bound on the constant $C$
\be
C \, < \, (g^*)^{1/2} \left( \frac{T_{eq}}{T_c} \right)^{1/2} \left( \frac{T_c}{T_m} \right)^2 \, .
\ee
In order for $\varphi$ to yield a successful candidate for Dark Matter, the temperature $T_m$ needs to be higher than $T_{eq}$. This condition means that the $\chi$ field must start out with less energy than would be the case in perfect thermal equilibrium of all fields.

The value, $M$, of the mass of dark-matter particles must then be chosen sufficiently small such that condition (\ref{cond2}) is satisfied. This yields a lower bound
\be
M \, < \,  \, (g^*)^{1/2} \left( \frac{T_{eq}}{T_c} \right)^{1/2} \frac{T_c^2}{f} \, \sim \, \mathcal{O}(10^{-28})  {\rm{eV}}  \, .
\ee
This value is  too low to satisfy the best observational constraints on the mass of an ultralight dark matter particle \cite{bounds}. However, this problem may be solvable by a slight modification of our model.
 
Without more detailed understanding of the intial conditions of the cosmological evolution,
we are not able to quantitatively determine the contribution of the fluctuations of $\varphi$ to the energy density of
Dark Matter. However, we argue that it is quite reasonable to expect that the fluctuations of $\varphi$ at times before the phase transition dominate the energy density of Dark Matter. If the mass of dark-matter particles is chosen as in \eqref{DM-mass}, 
the amplitude, ${\cal A}$, of $\varphi$-oscillations sufficient to obtain the right dark-matter energy density
at the time $t_{eq}$ of equal matter and radiation (corresponding to a temperature
$T_{eq} \approx 10\, {\rm eV}$) is estimated to be
\be
{\cal A} \, \sim \, T_{eq} \frac{T_{eq}}{m_{DM}} \, \sim \, 10^{21} {\rm GeV} \, .
\ee
Potentially, fluctuations of the $\chi$-field could also contribute to Dark Matter at early times.
However, as long as, at the initial time, the contribution of such fluctuations to the energy density
is smaller than the contribution of visible matter to the energy density,
it will never dominate at later times. 

\section{Remarks on Baryogenesis}

We first address the problem of \textit{baryogenesis}; see \cite{Cohen, Kamada}.
We assume that $\varphi$ is
coupled to the baryon current, $j^{\mu}_{B}$, via an interaction term in the Lagrangian of the form
\be \label{deltaL}
\delta {\cal L} \, = \, \frac{\alpha}{f} \partial_{\mu} \varphi\, \,j_B^{\mu}\, ,
\ee
where $\alpha$ is a dimensionless coupling constant.\footnote{$\alpha$ could be proportional
to the expectation value of a pseudo-scalar axion field.} 
Upon integrating by parts in the action functional and using the chiral anomaly to rewrite the divergence of the baryon current we find that this term is equivalent to \cite{Trodden}
\begin{equation}\label{anomaly}
\delta \mathcal{L}= \frac{\alpha}{f}\cdot\frac{3 g^{2}}{64\pi^2} \varphi\,\varepsilon^{\mu\nu\rho\sigma} F^{Y}_{\mu\nu} F^{Y}_{\rho\sigma}\,
\end{equation}
(we only include the contribution from $U(1)_Y$ in eq. (\ref{anomaly}) for illustrative purposes).
Here $g$ is a gauge coupling constant, and $F^{Y}$ is the hypermagnetic field strength.
As discussed in \cite{Cohen} under the term {\it ``spontaneous baryogenesis''}, the term $\delta \mathcal{L}$ introduced in \eqref{deltaL} can give rise to a non-vanishing baryon number density.\footnote{Note that Sakharov's criteria for
baryogenesis \cite{Sakharov} are satisfied in our model: The term $\delta \mathcal{L}$ in \eqref{deltaL} leads to
baryon-number violating processes. Given the uni-directional motion of $\varphi$ ($\dot{\varphi}>0$), this term favors
 a $CP$- and $C$-asymmetric state of the Universe. Furthermore, the field $\varphi$
is not in thermal equilibrium.} Assuming that $\varphi$ is homogeneous
in space,\footnote{A homogeneous initial state at early times might be the result of
inflation. However, all that really matters is that, in the patch that becomes the visible part
of the Universe, $\varphi$ has a negative homogeneous initial value,
so that it will slide down the potential hill towards the local minimum of its effective
potential.} the time derivative, $\dot{\varphi}$, of $\varphi$ is proportional to a chemical potential
conjugate to baryon number whose magnitude is given by 
\be
\mu_B \, = \, \frac{\alpha}{f} \,\,\dot{\varphi} \, ,
\ee
(see \cite{Cohen}).
As long as the baryon-number violating interactions involve degrees of freedom in thermal equilibrium,
a chemical potential $\mu_{B}$ yields a baryon number density, $n_B$, of the
order of
\be
n_B \, \sim \, \mu_B T^2 \, ,
\ee
and the induced baryon-number density-to-entropy ratio becomes
\be
\frac{n_B}{s} \, \sim \, \frac{\alpha}{f} \frac{1}{T_B} \dot{\varphi}(t_B) \, ,
\ee
where $T_B$ is the temperature when baryon-number violating processes
involve degrees of freedom that are no longer in equilibrium, and $t_B$ is 
the corresponding time. 

Dimensional analysis suggests that
\be \label{fieldvel}
\dot{\varphi} \, \sim \, \frac{f}{t} \left( \frac{1}{ft} \right)^p \, ,
\ee
where $p$ is a constant. One way to see this is to consider the equation of motion for $\varphi$ 
\be
{\ddot{\varphi}} + 3 H {\dot{\varphi}} \, = \, \frac{M^2}{f} (\varphi^2 - 2 \varphi f ) e^{-\varphi/f} \, .
\ee
If the source term is negligible, then we immediately obtain ${\dot{\varphi}} \sim a(t)^3$  and hence $p = 1/2$ (in the radiation dominated period). On the other hand, if the source term is dominant, then we can find an approximate solution in the field region where $\varphi < -f$ .
Setting $\varphi^2 =\text{ const. } \approx f^2$ on the right side, one arrives at an equation with solutions
growing logarithmically in time, for which (\ref{fieldvel}) holds with $p = 0$.

Relating time to temperature (using the Friedmann- and the Stefan-Boltzmann equations), one finds that 
\be \label{BGresult}
\frac{n_B}{s} \, \sim \, \alpha \left( \frac{T_B}{m_{pl}} \right)^{1 + 2p} \, .
\ee
It should be noted that the fact that the sign of $\dot{\varphi}$ is constant at early times, 
i.e., that the motion of $\varphi$ is uni-directional, which plays an important role in the above arguments, 
arises naturally in our scenario. Inserting the Planck mass and the value, $T_B$, of the temperature at which
sphalerons freeze out, we obtain a value for the baryon-to-entropy ratio that is small (unless the
coupling constant $\alpha$ is large).

Our estimate of the net baryon number is based on the assumption
that the initial value of $\varphi$ is negative throughout the region of space that will become 
the visible Universe. 
Once the degrees of freedom involved in baryon-number violating processes are no 
longer in equilibrium, which happens after the time, $t_{EW}$, of electroweak symmetry breaking,
a change in the chemical potential, $\mu_{B}$, conjugate to the baryon number does not affect the baryon asymmetry
that has arisen earlier. Given the parameters of our model, tuned in such a way
that $\varphi$ describes Dark Energy at late times, it is natural
to assume that the growth of $\varphi$ towards the value corresponding to the 
minimum of its effective potential continues until after the time of electroweak symmetry breaking.
 Thus the rate of chnage, $\dot{\varphi}$, 
of the field $\varphi$ can be expected to be strictly positive at time $t_{EW}$. The baryon-to-entropy ratio is then
determined by the value of $\dot{\varphi}$ at a time approximately given by $t_{EW}$.
This is what we have been using in our calculations. We repeat that the
relaxation of the baryon number density stops once the fermions have become massive \cite{Cohen}.

\section{Conclusions and Discussion} \label{conclusion}

In this paper we have presented an explicit realization of a cosmological scenario, 
originally proposed in \cite{us}, in which a single scalar field $\varphi$ is a source of Dark Matter
\textit{and} of Dark Energy. The cosmological evolution predicted by the model studied in this paper 
involves a phase transition at a late time $t_c$ corresponding to a
critical temperature $T_c$, which is driven by a second scalar field $\chi$. At temperatures $T>T_c$, 
(but $T$ well below the temperature of the electro-weak transition), the effective
potential traps $\varphi$ near a minimum, and oscillations about this
minimum yield a gas of cold dark-matter particles. After the $\chi$-phase
transition, the local minimum in the effective potential of $\varphi$ disappears, allowing a homogeneous
component of $\varphi$ to slowly roll and hence to become a natural candidate for Dark Energy.
If the transition temperature $T_c$ is below the temperature where structures
on galaxy-cluster scales freeze out from the expansion of the cosmological
background, the $\chi$-phase transition will not take place inside
non-linear structures, and a gas of dark-matter particles continues to prevail in
these regions.

The change in the role played by the field $\varphi$ 
-- from being a source of Dark Matter to being a source of Dark Energy -- corresponds
to the disappearance of a minimum in the effective potential of $\varphi$ at late times
and is driven by coupling $\varphi$ to a second scalar field $\chi$, which undergoes a
symmetry-breaking phase transition at the temperature $T_c$. 
We have also argued that, when the gradient $\partial_{\mu} \varphi$ is coupled to the baryon current, 
the dynamics of the field $\varphi$ during an early stage in the evolution of the Universe
 may yield a mechanism responsible for spontaneous baryogenesis.

There are observational constraints on models which involve interacting Dark Matter and Dark Energy (see e.g. \cite{Amendola}). If the change in the energy density of Dark Matter obeys the equation
\be
{\dot{\rho}_m} \, = \, - 3 H \rho_m - C \rho_m {\dot{\varphi_0}} \, ,
\ee
where $\varphi_0$ is the Dark Energy field and $C$ is a coupling constant with units of inverse mass, then the constraint on $\beta = C / m_{pl}$ is $\beta \lesssim 0.1$. In our case, $\beta$ depends on space. Inside of overdense regions we have $\beta = 0$ because there is no dynamical Dark Energy field. In the bulk, the background value of $\varphi$ becomes dynamical, and there is a coupling between Dark Energy and the residual Dark Matter fluctuations since the mass of the Dark Matter fluctuations depends on time. We get $\beta \sim m_{pl} / f$ (up to factors of order 1). Hence, the observational constaints are not a problem.

It would be an interesting problem to work out the constraints from observations on a model with the kind of space-dependent $\beta$ which we have. But this goes far beyond the scope of our current paper.

We have not addressed the connection between our 
model and fundamental physics. It is clear that the
Lagrangian (\ref{bare}) must be viewed as an effective
Lagrangian to be derived from a more fundamental theory. Insights 
in this direction will be described in a future paper.

The model studied in this paper does not solve the cosmological constant problem,
nor does it shed light on the Dark Energy coincidence problem. The latter is reflected in
some of the conditions (see e.g. (\ref{cond6})) we have to impose on the parameters of
our model. However, since the model involves dumping Dark Matter into
Dark Energy at a late time corresponding to the temperature $T_c$, it
may alleviate the ``Hubble tension'' which the standard
$\Lambda$CDM model suffers from.

\section*{Acknowledgement}
\noindent RB thanks the Institute for Theoretical Studies at the ETH Zuerich for hospitality
during early stages of this project, and the Institute for Theoretical Physics of the
ETH for hospitality at later stages. RN would like to thank Takashi Toma for useful
discussions. The research at McGill is supported in part by funds from NSERC and from the Canada 
Research Chair program.

\end{document}